# Comparison of SVM and Spectral Embedding in Promoter Biobricks' Categorizing and Clustering

Shangjie Zou
South China Agricultural University, College of Life Sciences
Guangzhou Institute of Advanced Technology, CAS

Abstract:

*Background:* In organisms' genomes, promoters are short DNA sequences on the upstream of structural genes, with the function of controlling genes' transcription. Promoters can be roughly divided into two classes: constitutive promoters and inducible promoters. Promoters with clear functional annotations are practical synthetic biology biobricks. Many statistical and machine learning methods have been introduced to predict the functions of candidate promoters. Spectral Eigenmap has been proved to be an effective clustering method to classify biobricks, while support vector machine (SVM) is a powerful machine learning algorithm, especially when dataset is small.

*Methods:* The two algorithms: spectral embedding and SVM are applied to the same dataset with 375 prokaryotic promoters. For spectral embedding, a Laplacian matrix is built with edit distance, followed by K-Means Clustering. The sequences are represented by numeric vector to serve as dataset for SVM trainning.

*Results:* SVM achieved a high predicting accuracy of 93.07% in 10-fold cross validation for classification of promoters' transcriptional functions. Laplacian eigenmap (spectral embedding) based on editing distance may not be capable for extracting discriminative features for this task.

Availability: Codes , datasets and some important matrices are available on github
https://github.com/shangjieZou/Promoter-transcriptional-predictor/tree/source-code

## INTRODUCTION

Synthetic biology is a promising research field, for it is providing solutions to many industrial problems. The basic principle of industrial synthetic biology is creating engineered living system by assembling DNA sequences [1]. These DNA sequences are named as DNA parts or "biobricks". iGEM (Internatonal Genetically Engineered Machine), a competition established by MIT, has a database called 'Registry of Standard Biological Parts' (http://parts.igem.org/), which is collecting information about thousands of standardized biobricks. Biobricks recorded in this database are mostly tested by former iGEM participants [2].

Promoter is a important category of biobrick, which controls the initiation of transcription. Promoters can be roughly divided into two classes: constitutive promoters, whose activity is not affected by any transcription factors, and inducible promoters, whose activity can be positively or negatively regulated by transcription factors. By choosing different promoters, we can tailor the expression of target genes in genetic circuits [3].

The aim of this study is evaluating and comparing SVM and spectral embedding's performances in categorizing promoters based on promoters' transcriptional functions (whether it can be regulated by transcriptional factors). Two clustering and classifying methods has been conducted on the same dataset. Laplacian eigenmap (spectral embedding) is a nonlinear dimensionality reduction method. It has been proved to be effective in discriminating biobricks and assessing biology databases' quality [4]. In this work, edit distance is used to represent the distance between promoters' sequences and generate Laplacian matrix. Through eigendecomposition, the eigenvectors and eigenvalues of Laplacian matrix are extracted for clustering. SVM is a powerful machine learning method, which can conduct nonlinear classification by applying kernel function to map samples in higher dimensional space and divide them with decision surface[5]. In the past years, SVM has been introduced to bioinformatics research and provided great performance in many applications, like disease diagnosis [6] and prediction of DNA parts' function [7].

## METHODS AND DATASET

*Computational Platform and tools*
The project is conducted on based on python 3.6 with Microsoft Windows 10 operating system. Edit Distances of promoter sequences are calculated by dynamic programming. The matrix of edit distances is used to generate Laplacian Matrix, followed by eigendecomposition and K-Means clustering. Before training SVM model, the sequences of promoters are translated to digital vectors through one-hot strategy: for the four deoxynucleosides, "A" is represented by [1,0,0,0], "G" is [0,1,0,0], "C" is equal to [0,0,1,0] and "T" is equal to [0,0,0,1]. SVM and K-Means are implemented by python module sci-kit learn.

*Dataset*
The Dataset is obtained from "Registry of Standard Biological Parts": http://parts.igem.org/Promoters/Catalog. 441 E.coli $\sigma^{70}$ promoters are obtained by web crawler as the "whole dataset". After filtering sequences larger than 500 bp, there are 375 sequences left in "filtered dataset". There are three categories of promoters: constitutive promoters, positively regulated promoters and repressible promoters. Both positively regulated promoters and repressible promoters are inducible promoters. In this study, two methods are used to split the dataset: Hold-Out and K-Fold. Hold-Out randomly divide the dataset into a trainning set (with 80% of the sequences) and a test set (with 20% of the sequences). Hold-Out is quick and simple, so it is used to divide dataset for tuning parameters. However, Hold-Out is not suitable for validation, so K-Fold method is introduced. This method would divide the dataset into K subsets. There would be K times of validation, and in each time, one subset would perform as test set, while the others are used for training.

*Laplacian Matrix and Spectral Clustering*
In this study, edit distance is used to represent the distance between two sequences. The edit distance *d(xi, xj)* is the minimum summation of edit operations (including deletion, insertion and substitution) that can transform a string *xi* to string *xj*. Dynamic programming is the most widely used algorithm for calculation of edit distance.

The matrix of edit distance is then normalized by equation 1, so as to generate a matrix of normalized edit distance (denoted as matrix M):

$$Mij = nomalized\_d(xi,xj) = \frac{d(xi,xj)}{\max(length(xi), length(xj))} \quad (1)$$

The next step is constructing similarity matrix S. Refering to former study [4], I adopted Gaussian kernel with $\sigma$ =0.3. The matrix S is calculated by equation 2:

$$Sij = e^{-\frac{Mij^2}{2\sigma^2}} \quad (2)$$

A diagonal matrix D is also needed to obtain Laplacian Matrix [8]. The diagonal of matrix D is the summation of similarities on the corresponding row:

$$Di = \sum_{1 \leq j \leq n} Sij \quad (3)$$

Finally, the Laplacian matrix G is obtained by: G = D - S. Eigendecomposition is applied on matrix G. The eigenvectors of matrix G are fed to K-Means clustering for classifying and visulization.

*Support Vector Machine (SVM)*
SVM is implemented with sci-kit learn. The dataset is filtered dataset, with 375 promoters whose lengths are shorter than 500bp. The dataset consists of 83 constitutive promoters and 292 inducible promoters. Before performance measurement, the dataset is randomly split into a training set (80%) and a test set (20%). In this study, parameter C and gamma are hyperparameters that should be tuned. So as to find the optimal combination of parameters, a nested loop is employed. Fig 1 shows the accuracy that different combinations of C and gamma achieved on test set. Based on the results, the model can achieve the highest prediction accuracy of 93.33% on test set when C is set to 10 and gamma is set to 0.015. The quick reduction of accuracy corresponds to increase of C and gamma might be a symbol of overfitting.

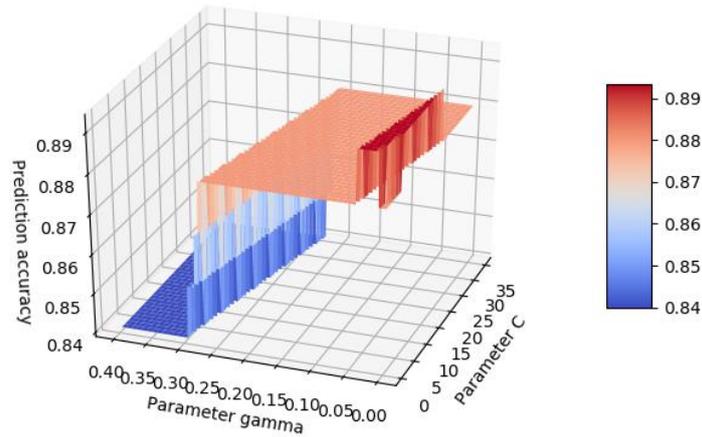

Fig 1. Optimizing parameter C and gamma

# RESULTS

*Spectral Embedding*

To comprehensively assess the performance of spectral embedding, two steps of measurements are conducted. Initially, the matrix of edit distance was built among the 441 σ70 promoters' sequences obtained from database (with 90 constitutive and 351 inducible). According to the clustering result (Fig 1), the 441 promoters can be distributed to two clusters. However, the two clusters are not really reflecting the categories of promoters' transcriptional functions. The bigger cluster (cluster_0) is containing 92.22% of constitutive promoters and 85.14% of inducible promoters.

This phenomenon, according to my hypothesis, is caused by the huge effect that the sequential lengths impose on edit distances. The distribution of lengths for sequences in these two clusters (Fig 2) can reflect this hypothesis.

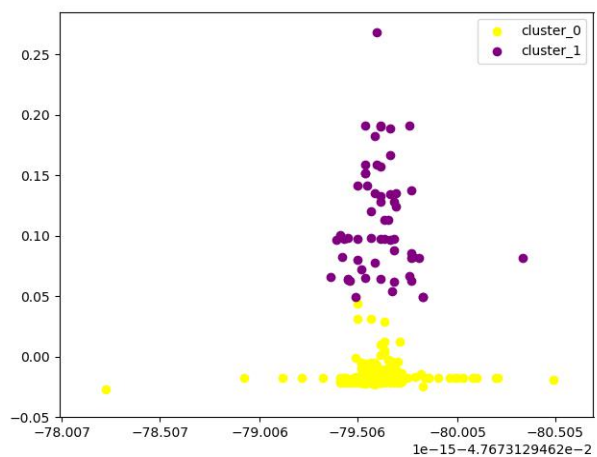

Fig 2. Spectral clustering result on all 441 sequences

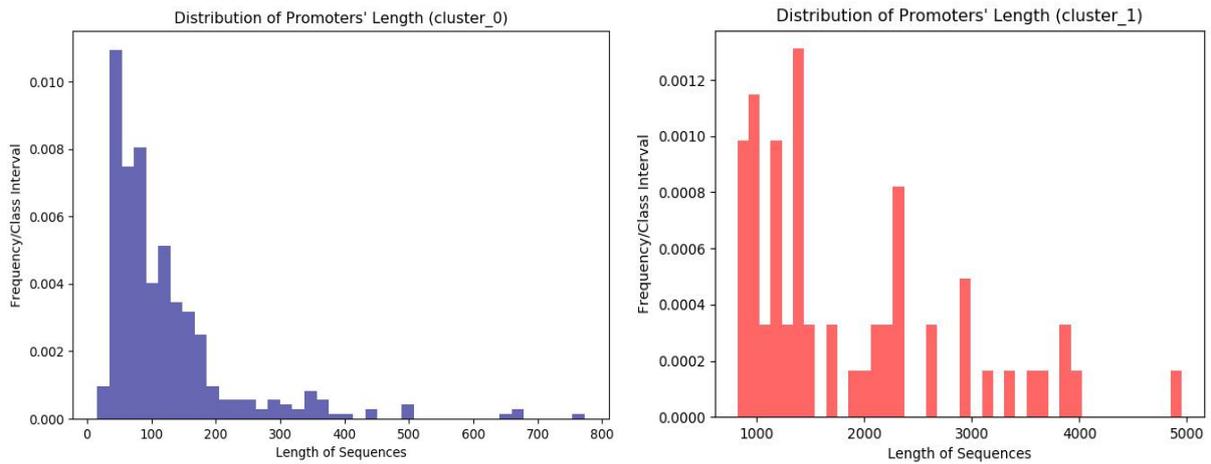

Fig 3. Distribution of lengths for sequences in two clusters

After re-checking the database, I found that most of the sequences that are distributed to cluster_1 are not actually promoters. For example, BBa_K256018 is labeled as constitutive promoter, but is actually a comprehensive gene circuit with coding gene, ribosome binding set and terminator. To avoid the potential effects of these false-labeled promoters, I filtered those promoters whose sizes are larger than 500 bp and redid clustering. However, based on the clustering result (Fig 3), the two clusters are still discriminated based on one eigenvector, which has been proved to be representing the lengths of sequences. Based on these results, spectral embedding based on edit distance may not be suitable for this task, because sequential length outweighs the other discriminative features.

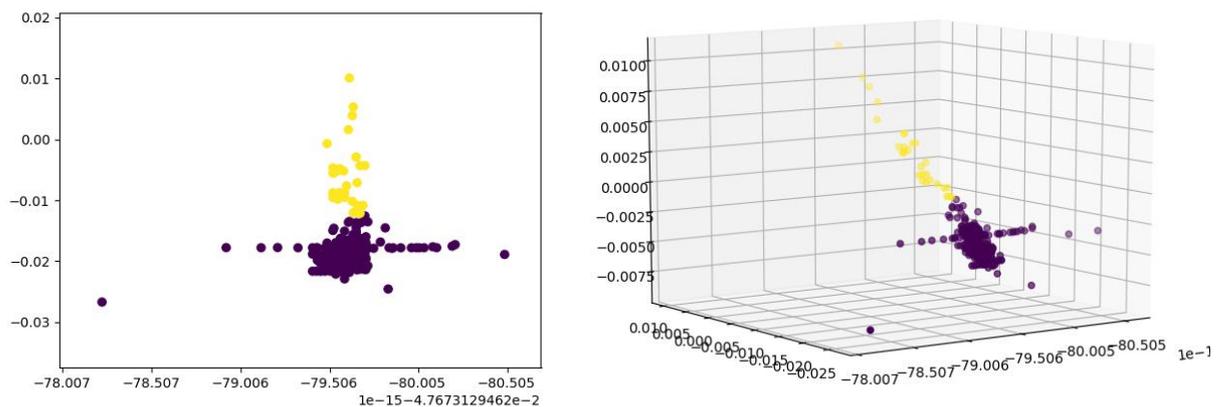

Fig 4. Clustering result of 375 promoters (filtered sequences larger than 500bp). Each axis is representing an eigenvector.

*Support Vector Machine (SVM)*
SVM is applied on the filtered dataset (375 promoters which are shorter than 500bp). Before training the model, the optimal parameters are tuned on training set and test set divided by hold-out method (refer to section METHODS AND DATASET). The SVM model with the best combination of parameters (C=10, gamma=0.015) is measured by cross-validation. In this study, the cross validation is conducted from 2-fold to 10-fold. According to fig 5, the model achieved the highest average accuracy of 93.07% when validated with 10-fold. It's obvious that 10-fold cross validation can provide the model with larger trainning sets, which may benefit its predicting performance on test set.

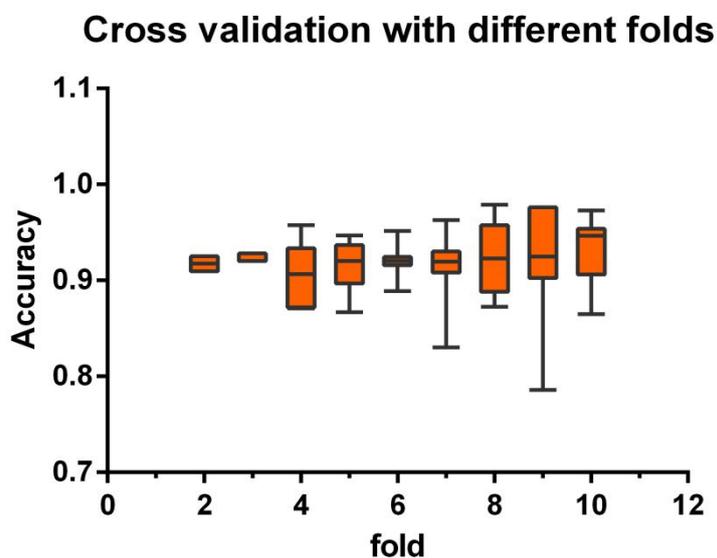

Fig 5. Cross validation score of optimized SVM model

## DISCUSSION

Based on the results, SVM is a powerful method for promoters' functional classification. Spectral embedding on the basis of edit distance is not performing well in this task.

The biggest shortcoming of the spectral embedding model that led to its poor performance is the huge effects of sequential lengths. When input sequences are vary, the length of sequences may outweigh other discriminative features. As a consequence, the clusters are divided based on the sequential lengths, without considering other features.

What's more, it's well known that different deoxyribonucleotide are having different physicochemical properties, which may contribute to promoters' performance. For example, guanine (G) and cytosine (C) can make the DNA fragment more stable, while those fragments with more adenine (A) and thymine (T) are more likely to unwind and initialize transcription. However, the algorithm of edit distance is not giving weighs based on deoxyribonucleotides' properties. Consequently, many important information may have lost when generating Laplacian matrix.